\documentclass[prl,showpacs,preprintnumbers,amsmath,amssymb,tightenlines,twocolumn,superscriptaddress]{revtex4}

\usepackage{graphicx}% Include figure files
\usepackage{dcolumn}% Align table columns on decimal point
\usepackage{bm}% bold math
\usepackage{epsfig}
\usepackage{stmaryrd}
\usepackage{color}

\newcommand{\bra}[1]{\langle\,{#1}\, |}
\newcommand{\ket}[1]{|\,{#1}\,\rangle}

\newcommand{\lrb}[1]{\langle\, {#1}\,\rangle}

\newcommand{\Real}{\mbox{Re}}

\newcommand{\fref}[1]{Fig.~\ref{#1}}

\newcommand{\eref}[1]{Eq.~(\ref{#1})}

\newcommand{\cref}[1]{chapter~\ref{#1}}

\newcommand{\Cref}[1]{Chapter~\ref{#1}}

\usepackage{ulem}  %unter- (\uline{important}) und durchstreichen (\sout{wrong})
\begin{document}

\title{On the Equivalence of Quantum and Classical Coherence in Electronic Energy Transfer}

\author{John~S.~Briggs}
\affiliation{Max Planck Institute for the Physics of Complex Systems,
N\"othnitzer Strasse 38, 01187 Dresden, Germany}

\email{briggs@physik.uni-freiburg.de}
\author{Alexander Eisfeld}
\affiliation{Max Planck Institute for the Physics of Complex Systems,
N\"othnitzer Strasse 38, 01187 Dresden, Germany}
\email{eisfeld@pks.mpg.de}
\begin{abstract}
To investigate the effect of quantum coherence on electronic
energy transfer, which is the subject of current interest in photosynthesis,
we solve the problem of transport for the simplest  model of an aggregate of
monomers interacting through dipole-dipole forces using both quantum and
classical dynamics. We conclude that for realistic coupling strengths quantum
and classical coherent transport are identical.
This is demonstrated by numerical calculations for a linear chain and for the photosynthetic Fenna-Matthews-Olson (FMO) complex 
\end{abstract}
\maketitle

\section{Introduction}
The role of electronic excitation transfer (EET) in the photosynthetic process
has been emphasised since the 1938 study by Franck and Teller
\cite{FrTe38_861_}. Already in that paper EET was suggested to occur through
the coherent quantum motion of  Frenkel exciton waves \cite{Fr31_17_}. Recently
renewed interest \cite{PlHu08_113019_,RoEiWo09_058301_,CoSc09_369_,IsFl09_234111_,PaHaKe10_12766_,WiMcMi09_10.1098,AbMu10_064510_,HoSaWh10_065041_} has been awakened by spectroscopic observations pointing to long-lived quantum coherences in the process of energy transfer in photosynthesis \cite{EnCaRe07_782_}. 
In particular it has been claimed that the speed of EET
is enhanced by quantum effects \cite{HoSaWh10_065041_}, which themselves are manifestations of quantum entanglement. The analogy has been made to advantages in quantum over classical information processing.
Considerable publicity has been accorded these claims which have been suggested to be  the first example of 'quantum weirdness in physiology' \cite{Wo09_17247_}.
Hence it appears timely to test this hypothesis on the simplest exciton model of 
an aggregate of dipole-coupled monomers, which we treat by both quantum and
classical mechanics. We demonstrate, both numerically and analytically, that
for dipole coupling strengths relevant to EET in molecular aggregates,
the coherences in quantum transport (from the Schr\"odinger equation) are
identical to those occurring in classical transport according to Newton's
equation. Correspondingly, in this fundamental example, the
effect on the EET mechanism arising due to quantum entanglement  is  mirrored  by the coherence of classical dipole-dipole EET. 

Although coupling to internal nuclear degrees of freedom and to the
environment plays an important role in EET (see e.g.\ Refs.\
\cite{RoEiWo09_058301_,FrRaeTi09_102_,HeKlBa02_1_,WAtEi10_053004_,PlHu08_113019_,ReMa98_4381_,RoScEi09_044909_,MueMaAd07_16862_,MoReLl08_174106_}),
here we will concentrate on the most basic situation in which quantum
coherence appears. 
Accordingly we consider  the idealised case of an aggregate of monomers,
each having only one electronic excited state, whose  purely electronic
degrees of freedom are coupled by dipole forces, shorn of any complications
arising from dissipation or decoherence due to vibronic or environment
coupling.
Quantum transport will be determined
by solving the time-dependent Schr\"odinger equation (TDSE). 

To treat the quantum system classically we must relate the classical motion to quantum properties. Here we  consider the monomers to consist of an oscillating
classical electron whose dipole strength will be set proportional to the
quantum oscillator strength. Exactly the model we use here has also been used
to discuss excitonic 
light absorption  by solids and liquids (see e.g.\ Refs.~\cite{Fa60_451_,McBa64_581_,NiRi67_4445_}) and  energy transfer on dimers \cite {Mc64_409_,Ku70_101_}
and nanoparticle arrays \cite{BrHaAt00_R16356_,MaMaKn08_2369_}. 
\\

%%%%%%%%%%%%%%%%%%%%%%%%%%%%%%%%%%%%%%%%%%%%%%%%%%%%%%%%%%%%%%%%%%%%%%%%%%55
\section{Quantum Mechanics}
%%%%%%%%%%%%%%%%%%%%%%%%%%%%%%%%%%%%%%%%%%%%%%%%%%%%%%%%%%%%%%%%%%%%%%%%%%%%5

The time-dependent Schr\"odinger equation (TDSE) for a Frenkel exciton on an aggregate is
\begin{equation}
i\hbar \frac{\partial}{\partial{t}} \ket{\Psi(t)} =\big({\mathbf H_0} + {\mathbf V}\big) \ket{\Psi(t)}
\end{equation}
where $\mathbf H_0$ is the sum of the hamiltonians of non-interacting monomers and $\mathbf V$ is the total potential energy of the pairwise interactions between monomers.
Since we consider the propagation of a single electronic excitation along the aggregate, we expand the full time-dependent wavefunction as, 
\begin{equation}
\label{eq:Psi(t)1}
\ket{\Psi(t)} = \sum_m c_m(t) \ket{\pi_m},
\end{equation}
where  $\epsilon_n$ is the single-monomer transition energy and $\ket{\pi_n}$ denotes a state in which monomer $n$ is electronically excited and all other monomers are in their ground state. The full aggregate ground-state energy is set to zero.
The matrix elements of $\mathbf H_0$ are taken to be  
\begin{equation}
\label{eq:H_0}
\bra{\pi_n}\mathbf H_0\ket{\pi_m}= \epsilon_n\, \delta_{nm}
\end{equation}
To connect with the classical model of oscillating dipoles, we will take the interaction  operator $\mathbf V$ to be the dipole-dipole interaction
\begin{equation}
\label{eq:def_Vnm}
V_{nm}= \bra{\pi_n} {\mathbf V}\ket{\pi_m}=\mu_n \hat{\varepsilon}_n.T_{nm}.\hat{\varepsilon}_m \mu_m,
\end{equation}
where $\mu_n$ and $\hat{\varepsilon}_n$ denote the magnitude and
orientation respectively of the transition dipole matrix element of monomer $n$ and
\begin{equation}
T_{mn}=(1 - 3\tilde{R}_{mn} \tilde{R}_{mn}) /R_{mn}^3,
\end{equation}
where $\hat{R}_{mn}$ is the direction of the vector separation, with length
$R_{mn}$, of monomers $m$ and $n$.
Substitution of \eref{eq:Psi(t)1} in the TDSE leads to a set of coupled equations of first order in the time derivative,
\begin{equation}
\label{eq:quantum}
i\dot{c}_n = \frac{\epsilon_n}{\hbar} c_n(t)  + \sum_m \frac{V_{nm}}{\hbar} c_m(t)
\end{equation}
These are the quantum equations.

\section{Classical mechanics}

In the classical case we consider that the coupled quantum transition dipoles are modelled by classical oscillators in the same geometry. The Hamilton equations of motion for linearly-interacting oscillators of mass $M_n$ and frequency $\omega_n$ are,
\begin{align}
\label{eq:classical_eq_motion_x}
\dot{x}_n=&\frac{p_n}{M_n}
\\
\label{eq:classical_eq_motion_p}
\dot{p}_n= &- M_n \omega_n^2x_n -\sum_m K_{nm} x_m 
\end{align}
where the $K_{nm}$ are coupling coefficients 
determined by
\begin{equation}
x_n(t) K_{nm} x_m(t)=\vec\mu^{\,\rm class}_n(t)\cdot T_{nm}\cdot \vec\mu^{\,\rm class}_m(t)
\end{equation}
and the r.h.s.\ has to be related to the quantum mechanical dipole-dipole interaction.

The treatment of quantum transition dipoles, corresponding to
electronic excitation and de-excitation, as classical electric dipoles
requires incorporating the monomer quantum properties into a classical
model. Here we follow Fano \cite{Fa60_451_} in defining classical dipoles
whose dipole moment corresponds to the quantum transition dipole moment.
To this end we first introduce the dimensionless quantum oscillator strength
\begin{equation}
\label{eq:f}
f_n=2 \frac{  m_e \epsilon_n}{(e\hbar)^2}\  \mu_n^2,
\end{equation}
where $m_e$ is the oscillator (electron) mass, $e$ is the electron charge and $\epsilon_n= \hbar \omega_n$  relates the transition energy to the monomer oscillation frequency.
\eref{eq:f} allows us to represent the classical dipole  moment in terms of the quantum transition
dipole moment $\mu_n$. 
That is we take the classical dipole moment of monomer $n$ to be given by
\begin{equation}
\label{eq:mu_class}
\vec\mu^{\,class}_n(t) = \hat\varepsilon_n  \sqrt{f_n} e\ x_n(t),
\end{equation}
where $\hat\varepsilon_n$ is the  direction of the transition dipole of the
$n$-th monomer. After identifying $M_n$ with $m_e$ the classical coupling strengths are then given by,
\begin{equation}
\label{eq:Kmn}
K_{nm} = \hat{\varepsilon}_n.T_{nm}.\hat{\varepsilon}_m\ (e^2 \sqrt{f_n
f_m})  m_e.\ 
\end{equation} 
These coupling coefficients can now be related to the quantum
mechanical $V_{nm}$, given in \eref{eq:def_Vnm}. By inserting \eref{eq:f} into \eref{eq:Kmn} on finds
\begin{align}
\frac{K_{nm}}{ m_e\sqrt{\omega_n \omega_m}}%= &\ \hat{\varepsilon}_m.T_{mn}.\hat{\varepsilon}_n \ |\mu_n| |\mu_m|\ \frac{\sqrt{\omega_m/ \omega_n}}{\hbar} \\
= \frac{2V_{nm}}{\hbar},
\end{align}
To connect to the quantum equations, we 
introduce the dimensionless complex amplitude 
\begin{equation}
\tilde{z}_n(t)=\tilde{x}_n(t)+i \tilde{p}_n(t)
\end{equation}
with
\begin{align}
\tilde{x}_n&= \sqrt{\frac{M_n\omega_n}{2 \hbar}} x_n\\
\tilde{p}_n&= \frac{1}{\sqrt{2 \hbar M_n \omega_n}}p_n
\end{align} 
One then obtains
\begin{equation}
\label{eq:dot_z_final}
i\dot{\tilde{z}}_n= \omega_n \tilde{z}_n  + \sum_m \frac{ 2V_{nm}}{\hbar} \Real (\tilde{z}_m)
\end{equation}
This equation has to be compared to the quantum equation (\ref{eq:quantum}).
The difference is a factor of two and the appearance of $ \Real (\tilde{z}_m)$ instead of $\tilde{z}_m$ in the coupling terms.
The resulting difference in the dynamical equations is best seen by first taking the time derivative of Eq.~(\ref{eq:dot_z_final}) to obtain, using 
$
\tilde p_n = \dot{\tilde x}_n/\omega_n
$,
\begin{equation}
\label{eq:secorderz}
\ddot{\tilde z}_n =  - \omega_n^2 \tilde z_n - \omega_n \sum_m \frac{2V_{nm}}{\hbar} \tilde{z}_m - i\sum_m (\omega_m - \omega_n) \frac{2V_{nm}}{\hbar} \tilde{p}_m .
\end{equation}
By contrast, taking the time derivative of the TDSE Eq.~(\ref{eq:quantum}), we
obtain a second order equation for  the quantum amplitudes, 
\begin{equation}
\label{eq:secorderq}
\begin{split}
\ddot{ c}_n = & - \omega_n^2  c_n - \omega_n \sum_m \frac{2V_{nm}}{\hbar} c_m   - \sum_m (\omega_m - \omega_n)\frac{V_{nm}}{\hbar}  c_m \\ & -   \sum_{mm^\prime} \frac{V_{nm}}{\hbar}\frac{V_{mm^\prime}}{\hbar}  c_{m^\prime}.
\end{split}
\end{equation} 
Scaling the time as $\omega_nt$ this equation is,
\begin{equation}
\begin{split}
\ddot{ c}_n = & -  c_n -\sum_m \frac{2V_{nm}}{\epsilon_n } c_m - \sum_m (\frac{\epsilon_m}{ \epsilon_n} - 1)\frac{V_{nm}}{\epsilon_n}  c_m \\ & -   \sum_{mm^\prime} \frac{V_{nm}}{\epsilon_n}\frac{V_{mm^\prime}}{\epsilon_n}  c_{m^\prime}
\end{split}
\end{equation}
We note that the last term in this equation is of second order in the normalized coupling strength $V_{nm}/(\epsilon_n)$ and can be neglected when this coupling is small. This we call the realistic coupling approximation (RCA). The penultimate term is zero for identical monomers $\epsilon_m = \epsilon_n$ and small when the fluctuations in the transition energies are small relative to the mean. Under the same conditions the last term in the classical Eq.~(\ref{eq:secorderz}) is small.
In these two approximations, we can drop the last  two coupling terms in Eq.~(\ref{eq:secorderq}) and the last term in Eq.~(\ref{eq:secorderz}), so that these two equations become identical. To this order then the classical Eq.~(\ref{eq:dot_z_final}) and the quantum Eq.~(\ref{eq:quantum}) are equivalent and we can associate, up to a normalization factor, the quantum amplitudes $c_n$ with their classical counterparts $\tilde{z}_n$.
The classical "probability of occupation"  $P_n^{\rm\, class}(t)$ is then defined as
\begin{equation}
P_n^{\rm\, class}(t)\equiv\frac{ |\tilde{z}_n(t)|^2}{\sum_n |\tilde{z}_n(t)|^2}.
\end{equation}

In realistic cases of aggregates involving monomer molecules  one has the situation that the
interaction splitting is much less than the transition energy i.e. $ V_{nm} \ll \epsilon_n \quad \forall \quad n,m$, i.e. the RCA is valid.
Indeed, in the light harvesting complexes the electronic transition
energy of the chlorophyll monomers is $\sim 12\, 000$ cm$^{-1}$,
whilst the dipole-dipole interaction between the chlorophylls is in the order
of a few hundred cm$^{-1}$ \cite{AmVaGr00__}.
Hence we can conclude that all coherences of propagation in the
quantum case are also contained in the classical propagation for realistic
coupling strengths. Note that the quantum equation \eref{eq:quantum} can always be simulated by  classical oscillators coupled linearly in position (or dipole strength), so long as all
$(V_{nm}/\epsilon_n)$ are small and the relative fluctuations in the $\epsilon_n$ are small. We emphasize that this is quite different from a standard mapping \cite{Di27_243_,St66_36_} of electronic degrees of freedom to classical oscillators. This mapping gives an unrealistic complex classical Hamiltonian involving additional momentum couplings. Only in RCA do the equations (\ref{eq:dot_z_final}) for coupled classical electric dipoles  reproduce the quantum amplitudes of coupled transition dipoles i.e. only in this approximation is Eq.~(\ref{eq:dot_z_final}) a good approximation to Eq.~(\ref{eq:quantum}). This is the major result of this analysis.
\\

\section{Comparison of Quantum and Classical Mechanics}

In the following we will demonstrate the applicability of the realistic coupling equation with two examples.
The first one is a linear chain of identical monomers and identical coupling between neighboring monomers. Writing $V_{nm}=V$ and taking only nearest neighbor coupling into account the equation (\ref{eq:quantum}) reduces to
\begin{equation}
i\frac{dc_n}{d\tau} = \frac{1}{2} \big( c_{n-1}(\tau) + c_{n+1}(\tau)\big).
\end{equation}
Here the dimensionless time $\tau=(2 V/\hbar) t$ has been introduced.
One can show that the solution of these equations with initial condition $c_n(0) = \delta_{n0}$ are just the Bessel functions $ c_n = J_n(2Vt/\hbar)\exp(in\pi/2)$. 
Then, beginning on monomer zero, the excitation probability of monomer $n$ is
\begin{equation}
\label{eq:P_n_QM}
P_n(t) = J_n^2(2Vt/\hbar)
\end{equation}
This is the result obtained e.g.~in
Refs.~\cite{Me58_647_,RoScEi09_044909_}.  From the analytic result
\begin{equation}
\lrb{n^2} = \sum_n n^2 J_n^2(2Vt/\hbar) = (2V/\hbar)^2 t^2
\end{equation}
one infers a constant "velocity" of electronic propagation given by $
\sqrt{\lrb{n^2}}/t = 2V/\hbar$.  Hence in this simple model the transport of electronic energy is fully coherent. Note that, in nearest-neighbour coupling, the scaled time  $\hbar/(2V)$ is the natural time unit since it is just the mean time of EET from one monomer to its neighbours.  A detailed study of this coherent quantum transfer is given in Ref.~\cite{RoScEi09_044909_}.  Within the RCA the classical equations predict identical coherence and constant velocity of propagation.
%
%%%%%%%%%%%%%%%%%%%%%%%%%%%%%%%%%%%%%%%%%%%%%%%%%%%%%%%%%%%%%%%%%%%%%%%%%%
\begin{figure}[tp]
\includegraphics[width=9.5cm]{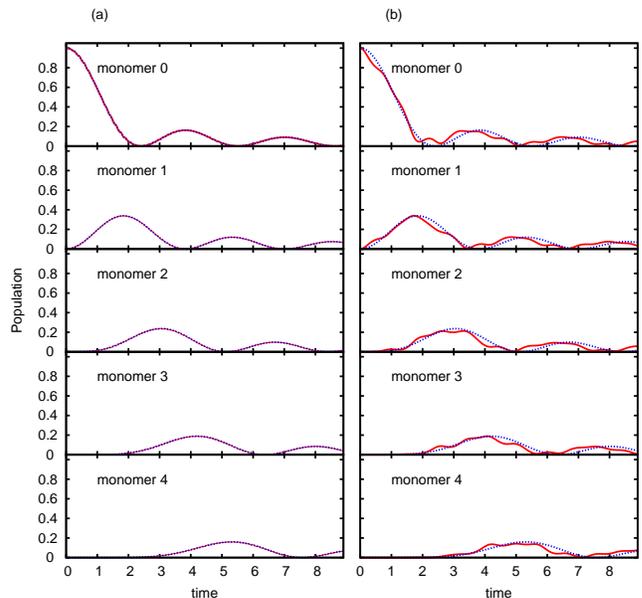}
\caption{\label{fig:classical}The probability as a function of time in units
of $\hbar/2 V$ of excitation of monomer $n$ for a linear chain. Blue: quantum result, red:
classical result. Left figures for $V/\epsilon= 1/40$  and right figures for $V/\epsilon=1/6$. } 
\end{figure}
%%%%%%%%%%%%%%%%%%%%%%%%%%%%%%%%%%%%%%%%%%%%%%%%%%%%%%%%%%%%%%%%%%%%%%%%%%
%
We have tested the validity of the RCA by performing exact numerical
evaluation of the full classical equations \eref{eq:classical_eq_motion_x} and \eref{eq:classical_eq_motion_p} for this case of identical monomers and nearest-neighbour interaction.
The $P_n^{\rm\, class}(\tau)$ for the classical energy propagation  along a chain of 19
monomers are compared to the quantum Bessel function prediction in
Fig.~\ref{fig:classical} where results for two different coupling strengths are shown. 
One sees {(Fig.~\ref{fig:classical}a) that for a small value $1/40$ of the parameter $V/\epsilon$,
nevertheless one that is typical for photosynthesis, the exact classical
probability that energy resides on a given monomer at a given time is indistinguishable from the
quantum Bessel function result.  Even for $V/\epsilon = 1/6$, which is
unrealistically large, one sees (Fig.~\ref{fig:classical}b)
that the classical result still follows closely the envelope of the
quantum result, although some deviations are beginning to appear.  Hence,
for realistic coupling, these numerical
results fully justify our analytic approximation leading to
identity of classical and quantum EET.

In the previous example of a linear chain all the nearest-neighbour coupling strengths and transition energies have been taken to be equal. 
To demonstrate that the classical and quantum results also coincide in more general cases, we have calculated energy transfer dynamics for an aggregate where the coupling elements and transition energies correspond to those usually assumed for the FMO complex \cite{AdRe06_2778_}. Coupling to the environment is neglected.
The complex consists of seven monomers and we have calculated the probability that excitation, starting localised on one monomer, has reached other monomers at later times.  
%The classical dynamics are given by solution of \eref{eq:classical}.
Typical results are shown in \fref{fig:FMO} for excitation initially localized on monomer 1. 
Since monomer 1 couples strongly to monomer 2 only the probabilities for these two monomers are of the same order. Also shown is the much lower probability of exciting monomer 3. The probabilities of excitation of other monomers are not shown but the agreement between quantum and RCA results is equally good.
The left column shows results using realistic FMO couplings and transition energies quoted in Ref.~\cite{AdRe06_2778_}.
One sees an oscillatory motion of excitation but most important, the classical and quantum results are indistinguishable and appear as a single curve.  
In the right column the transition energies have been artificially reduced by 12000 cm$^{-1}$ so that they become similar to the coupling energies. 
Then the RCA is no more valid and classical results (red) are not identical to quantum (blue), although even in this case the overall agreement is still very good.
Similar agreement has been found for all other monomers and also for different initial conditions.
\begin{figure}[pt]
\includegraphics[width=9cm]{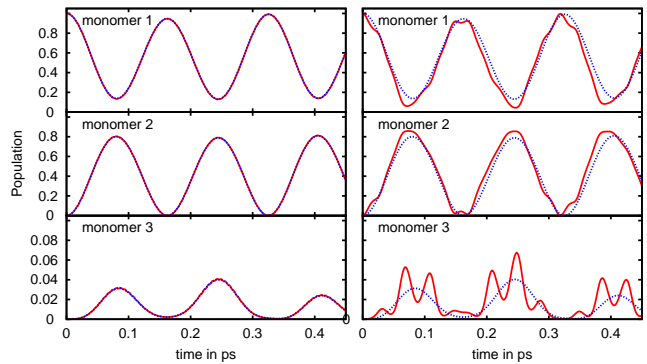}
\caption{\label{fig:FMO}Time dependent electronic occupation of monomer 1 (top), monomer 2 (middle) and monomer 3 (bottom) in the case of the FMO parameters. Initially the excitation is localised on monomer 1. The right column shows results where the transition energy is arbitrarily decreased by 12000 cm$^{-1}$.  }
\end{figure}

What is important here is not to infer that, since the quantum result agrees with the classical, there is no quantum entanglement. Rather it is  the classical amplitudes and phases  which mimic this entanglement. For the one-exciton manifold of pure states considered here, entanglement is expressed simply by the possibility to write the wavefunction, or parts of it, in a single product  form or not. A quantitative measure of the entanglement between sites $i$ and $j$ is the concurrence \cite{SaIsFl10_462_} which in the present situation is simply proportional to the matrix elements of the corresponding density matrix, i.e.\ $|\rho_{ij}| = |c_i^*c_j|$. As example, for the linear chain we have the analytic result $|\rho_{ij}(t)| = |J_i(2Vt/\hbar)J_j(2Vt/\hbar)|$ describing the development of entanglement in time. Note, however that the classical complex amplitudes will lead in RCA to exactly the same result. Indeed, quite generally we find that the agreement between quantum and RCA classical complex amplitudes, which determines the coherence, is equally close as for the population comparisons shown in the figures.

To summarise, the question as to whether quantum mechanics has an effect on EET on molecular aggregates appears from the results presented here to be answered with a definite 'no'.  One can conclude that the entanglement in the quantum case leads to no essential change from the  coherence characteristics of {\it{classical}}  electronic energy transfer.  The results presented above are obtained by neglect of the influence of an environment. We have also performed calculations including dephasing processes in the quantum and classical equations and found similar good agreement. Also, in a recent paper, Zimanyi and Silbey \cite{ZiSi10_144107_} have studied the related problem of energy transport between just two monomers of differing character in the presence of environmental effects. They show also in this case that a purely classical model leads to the same energy transfer probability as the quantum result. We conclude that some care must be
taken when attributing coherence  in energy transfer as due to quantum
mechanical effects.

%%%%%%%%%%%%%%%%%%%%%%%%%%%%%%%%%%%%%%%%%%%%%%%%%%%%%%%%%%%%%%%%%%%%%%%%%%%%%%
\section{Acknowledgement}
Thanks are due to Prof.\ H.\ Helm, Prof. W.\ T.\ Strunz and Prof.\ J.-M.\ Rost for their help, advice and insight.

%%%%%%%%%%%%%%%%%%%%%%%%%%%%%%%%%%%

%\bibliography{quantClassEET}

%%%%%%%%%%%%%%%%%%%%%%%%%%%%%%%%%%%
\end{document}